\documentclass{article}
\usepackage{spconf,amsmath,graphicx}
\usepackage{arydshln}
\usepackage{xcolor}
\usepackage{slashbox}
\usepackage{multirow}
\usepackage{amsfonts}
\usepackage{citesort}
\usepackage{booktabs}

\newcommand{\ra}[1]{\renewcommand{\arraystretch}{#1}}

\title{VarArray Meets t-SOT: Advancing the State of the Art of Streaming Distant Conversational Speech Recognition}
\name{Naoyuki Kanda, Jian Wu, Xiaofei Wang, Zhuo Chen, Jinyu Li, Takuya Yoshioka}
\address{Microsoft, One Microsoft Way, Redmond, WA, USA}
\begin{document}
\ninept
\maketitle
\begin{abstract}
\vspace{-.5em}
This paper presents a novel streaming automatic speech recognition (ASR) framework for multi-talker overlapping speech captured by a distant microphone array with an arbitrary geometry. Our framework, named t-SOT-VA, capitalizes on independently developed two recent technologies; array-geometry-agnostic continuous speech separation, or VarArray, and streaming multi-talker ASR based on token-level serialized output training (t-SOT). To combine the best of both technologies, we newly design a t-SOT-based ASR model that generates a serialized multi-talker transcription based on two separated speech signals from VarArray. We also propose a pre-training scheme for such an ASR model where we simulate VarArray's output signals based on monaural single-talker ASR training data. Conversation transcription experiments using the AMI meeting corpus show that the system based on the proposed framework significantly outperforms conventional ones. Our system achieves the state-of-the-art word error rates of 13.7\% and 15.5\% for the AMI development and evaluation sets, respectively, in the multiple-distant-microphone setting while retaining the streaming inference capability. 
\end{abstract}
\begin{keywords}
Multi-talker automatic speech recognition, conversation transcription, microphone array, streaming inference
\end{keywords}
\section{Introduction}
\label{sec:intro}
\vspace{-.5em}

Distant conversational speech recognition (DCSR), a task to 
transcribe multi-talker conversations captured by a distant microphone or a microphone array,
 has long been studied as one of the most challenging problems for automatic speech recognition (ASR) \cite{janin2003icsi,carletta2005ami,fiscus2007rich,watanabe2020chime,yu2022m2met}.
 Besides the acoustic distortion resulting from the long speaker-to-microphone distance,
 speech overlaps significantly degrade the ASR accuracy \cite{chen2020continuous,raj2020integration}
 while they often happen in natural conversations and are not negligible~\cite{ccetin2006analysis}. 
 The linguistic characteristics are also 
 complex due to frequent turn-takings. 
Given these difficulties, most studies
on DCSR
have been
conducted based on strong prerequisites
such as 
 the availability of utterance-level ground-truth segmentations (e.g., \cite{kanda2019guided,zhang2022bigssl}) 
or offline inference (e.g., \cite{kanda2019simultaneous,medennikov2020stc,ye2022royalflush}).
To advance the DCSR,
innovations in both front-end signal processing
and back-end ASR, as well as their efficient integration, would be needed.

 Continuous speech separation (CSS) is a front-end-based approach 
 to DCSR with streaming inference 
 \cite{yoshioka2018recognizing,chen2020continuous}.
Unlike traditional speech separation,
CSS converts a long-form multi-talker speech signal containing overlapping speech 
into multiple overlap-free speech signals with a fixed latency by using a sliding window. 
Each of the separated signals can then be passed to a conventional ASR system.
 VarArray, a recently proposed array-geometry-agnostic multi-channel CSS
model, 
showed impressive effectiveness 
in dealing with the speech overlaps in real meetings~\cite{yoshioka2022vararray}. 
VarArray can be applied to arbitrary microphone array geometries without retraining and thus enjoys a low adoption barrier. 
 However, the simple front-end and back-end concatenation leaves the back-end ASR system
 susceptible to the artifacts and errors caused by the CSS front-end.

On the other hand,
a great deal of effort has been 
made to extend the back-end ASR models to directly cope with the overlapping speech.
One approach is using a neural network with multiple output branches,
where each output branch generates a transcription for one speaker
(e.g., \cite{yu2017recognizing,seki2018purely,chang2019end,chang2019mimo,tripathi2020end,lu2021streaming,sklyar2021streaming}).
Another approach is 
serialized output training (SOT) \cite{kanda2020sot},
where an ASR model has only a single output branch
that generates multi-talker transcriptions one after another with a
special separator symbol.
Recently, a variant of SOT,
named token-level SOT (t-SOT), was proposed 
for streaming inference, which achieved 
 the state-of-the-art (SOTA) accuracies for several multi-talker ASR tasks \cite{kanda22arxiv}.
In the t-SOT framework, the ASR model generates 
recognition tokens (e.g. words or subwords) 
spoken by multiple overlapping speakers in chronological order with a special separator token.
These methods can produce highly accurate transcriptions by modeling the multi-talker multi-turn speech signals effectively in terms of both the acoustic and linguistic aspects \cite{kanda2021large}. 
However, 
most studies were conducted with monaural audio, and the existing multi-channel-based studies employed modules that are only applicable for offline inference 
\cite{chang2019mimo,zhang2021end,shen2022volcspeech}. 
 Also, less considerate multi-channel extensions of the ASR models could suffer from the
 high data transmission cost from the microphone array device 
 to the ASR server \cite{kanda2019acoustic}.

In this work,
we present a novel 
streamable DCSR framework which can leverage  arbitrary microphone arrays
with a limited data transmission cost.
Our framework is designed to achieve the best combination of VarArray and t-SOT and thus named t-SOT-VA (t-SOT with VarArray).
To this end, 
we newly design a t-SOT-based ASR model 
that takes in two separated speech signals from VarArray.
To help this ASR model correct the errors made by the front-end, 
 we also propose a pre-training scheme for 
 such ASR models based on simulated erroneous separated signals, where these signals can be simulated by using monaural single-talker ASR
training data. 
The VarArray front-end processing can be executed on the edge device, which allows the 
device to send only two audio streams to the server, possibly with lossy compression. 
These improvements enable highly accurate 
streaming DCSR with arbitrary microphone arrays.
We conduct a comprehensive evaluation of the proposed framework by using the AMI
meeting corpus \cite{carletta2005ami}.
Our system based on the proposed framework significantly outperforms various conventional
DCSR systems, achieving the SOTA word error rates (WERs) for the AMI test sets
while retaining the streaming inference capability.

\begin{figure}[t]
  \centering
  \includegraphics[width=1.0\linewidth]{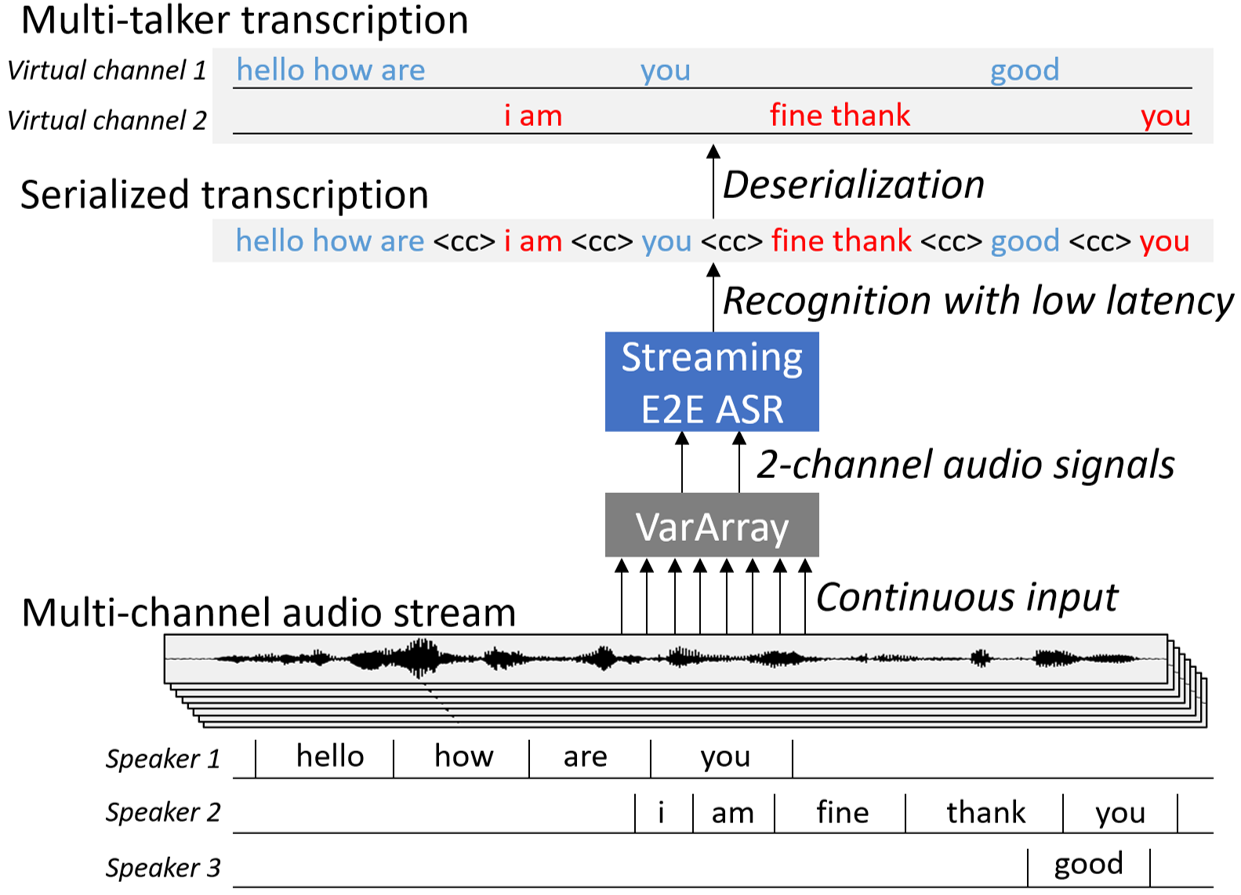}
  \vspace{-7.5mm}
      \caption{Proposed t-SOT-VA framework. }
  \label{fig:overview}
   \vspace{-6mm}
\end{figure}

\section{Related Works}
\vspace{-.5em}

\subsection{Array-geometry-agnostic CSS with VarArray}
\label{sec:vararray}
\vspace{-.5em}

VarArray is a recently proposed neural network-based CSS front-end that 
converts long-form continuous audio input
from a microphone array
into $K$ streams of overlap-free audio signals, where $K$ is usually set to two \cite{yoshioka2022vararray}. 
VarArray can cope with the input from
an arbitrary microphone array without retraining the model parameters.
This ``array-geometry-agnostic'' property is enabled by
interleaving temporal-modeling layers and geometry-agnostic cross-channel layers.
The model takes a multi-channel short-time Fourier transform as an input to 
output the time-frequency masks for $K$ speech sources and two noise sources, corresponding to stationary and transient noise sources. 
The estimated masks are used to form minimum variance distortion-less response (MVDR)
beamformers to produce $K$ overlap-free signals.
Streaming processing is realized based on a sliding window, where  
 $T^s$ sec of $K$ overlap-free
 signals are generated for every $T^s$ sec by using a $T^f$-sec look-ahead.
 In our experiment, both $T^s$ and $T^f$ were set to 0.4, thereby 
 causing the total algorithmic latency to be 0.8 sec.
 As a pre-processing step, we apply real-time dereverberation based on the weighted prediction error method~\cite{yoshioka2012wpe}. 
We refer the readers to \cite{yoshioka2022vararray} for more details.

\subsection{Streaming multi-talker ASR based on t-SOT}
\label{sec:t-sot}
\vspace{-.5em}

The t-SOT method was proposed to enable multi-talker overlapping speech recognition
with streaming processing \cite{kanda22arxiv}. 
With t-SOT,
only up to $M$ speakers can be active simultaneously. The following description assumes $M=2$ for simplicity. 
With t-SOT, 
the transcriptions for multiple speakers are 
serialized into a single sequence of 
recognition tokens (e.g., words or subwords) 
by sorting them in a chronological order. 
A special token $\langle \mathrm{cc}\rangle$, which indicates a change of `virtual' output
channels, 
is inserted between two adjacent words spoken by different speakers (such examples can be found in the middle of Fig. \ref{fig:overview}).
A streaming end-to-end ASR model \cite{li2022recent} is trained 
based on pairs of such serialized transcriptions and the corresponding audio samples.
During inference, 
an output sequence including 
$\langle \mathrm{cc}\rangle$ is generated
by the ASR model in a streaming fashion, 
which is then `deserialized' to generate separate transcriptions 
by switching between the two virtual output channels at each encounter with the $\langle \mathrm{cc}\rangle$ token (see the top of Fig. \ref{fig:overview}).
The t-SOT model outperformed prior multi-talker ASR models
even with streaming processing.
See \cite{kanda22arxiv} for more details.

\begin{figure}[t]
  \centering
  \hspace{10mm}
  \includegraphics[width=0.67\linewidth]{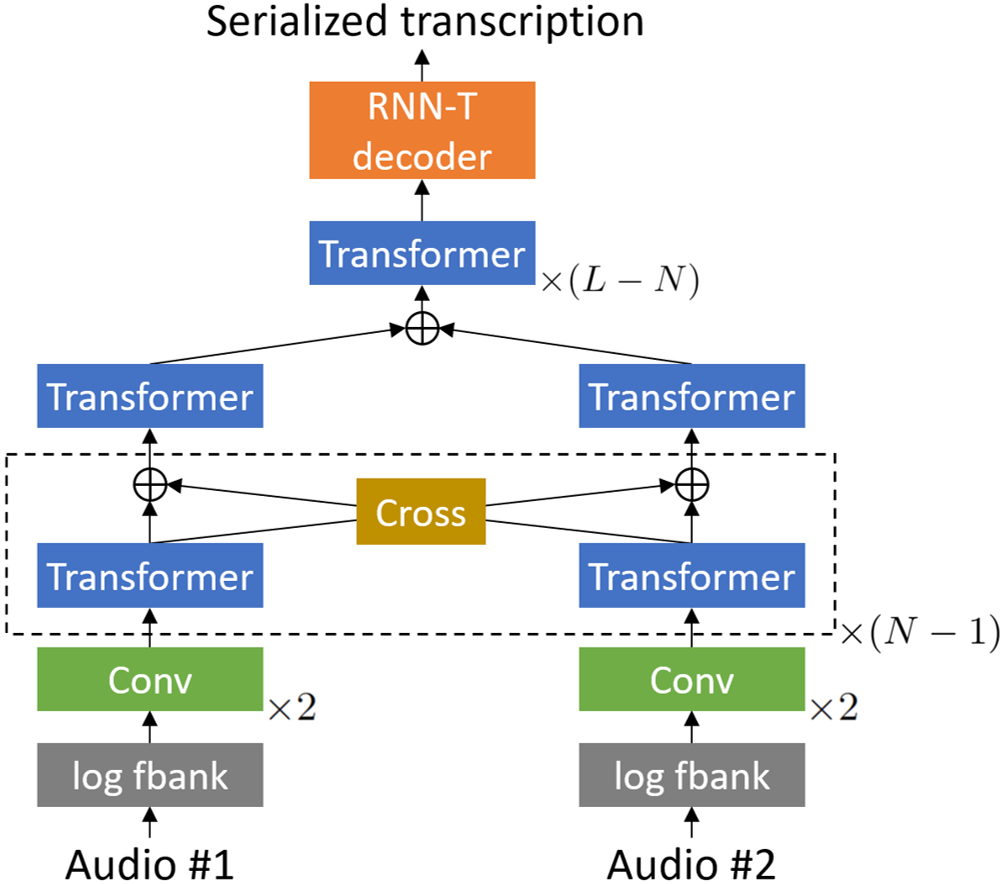}
  \vspace{-4.0mm}
      \caption{Proposed two-channel transformer transducer architecture.}
  \label{fig:asr}
   \vspace{-5.5mm}
\end{figure}

\section{t-SOT-VA: A New DCSR framework}
\label{sec:mc-t-sot}
\vspace{-.5em}

\subsection{Framework design: VarArray $\times$ t-SOT}
\label{sec:design}
\vspace{-.5em}

Fig. \ref{fig:overview} shows the overview of the proposed framework.
The input multi-channel audio signals are first
fed into the VarArray front-end, which generates two audio streams in a streaming fashion.
These audio streams are then provided to
the two-channel-input streaming end-to-end ASR model
to generate
the serialized transcription based on the t-SOT approach.
Finally, it is deserialized to produce
multi-talker transcriptions.

The proposed combination of VarArray and t-SOT ASR has multiple advantages. 
Firstly, thanks to the array-geometry-agnostic property of VarArray, 
the system is applicable to various microphone arrays without retraining.
Secondly, by executing the VarArray processing on the microphone array device,
we can limit the device-to-ASR-server data transmission needs 
to only two audio signals, 
which provides a large practical benefit.
Thirdly, 
training our ASR model is much simpler than other multi-channel  multi-talker ASR models
as the latter ones require simulating multi-channel training data with realistic phase information \cite{chang2019mimo,zhang2021end,shen2022volcspeech} while our model does not. 
Finally, our ASR model 
can be easily fine-tuned by using the VarArray outputs for real multi-channel recordings and the corresponding time-annotated reference transcriptions.
This is not the case with conventional systems that apply a single-talker ASR model to each separated signal. 
This is because, in order to fine-tune the single-talker model based on the separated signals, 
the reference transcriptions must be generated so that they match the separated signals, which can be tricky because the front-end may even split one utterance into different streams. 


\subsection{Two-channel transformer transducer}
\vspace{-.5em}

Fig. \ref{fig:asr} shows 
the proposed architecture of the two-channel 
 end-to-end ASR model. 
Our model is a modified version of the 
transformer transducer (TT) \cite{zhang2020transformer} and is obtained 
by splitting the transformer encoder to take in two audio signals.
First, the audio input of each channel
is converted to the log mel-filterbank, followed by  
two convolution layers.
The output from the last convolution layer is then fed into a stack of transformer layers
with cross-channel transformation. 
Let $x_{l,c}$ denote the output of the $l$-th transformer layer in the $c$-th
channel, where $c\in\{0,1\}$. 
The cross-channel transformation output, $\hat{x}_{l,c}$, is calculated in the following two different ways: 
\begin{align}
& \text{Scaling:}\;\;\;\;\; \;\;\;\;\; \;\;\:\,\hat{x}_{l,c} = x_{l,c} + s_l\cdot x_{l,(1-c)};~ \text{or} \label{eq:scale} \\
&\text{Projection:} \;\;\;\;\;   \hat{x}_{l,c} = x_{l,c} + \phi(W_l\cdot x_{l,(1-c)}+b_l), \label{eq:proj}
\end{align}
where $s_l$, $W_l$, and $b_l$ are a learnable scalar, a weight matrix, and a bias vector, respectively, for the $l$-th layer while 
$\phi()$ denotes a rectified linear unit activation function.
The outputs of the $N$-th transformer layer are summed across the channels, which is further processed by 
additional $(L-N)$ layers of transformer, with $L$ being the total number of the transformer encoders.
On top of the encoder, a recurrent neural network transducer (RNN-T)-based
decoder is applied to generate the serialized transcription described earlier. 

Note that the parameters for the two-branched encoders,
i.e.,  the parameters of the convolution layers, the first $N$ transformer layers, and the cross-channel
transformations, are shared among the two channels.
Therefore,
the parameter count of the two-channel TT is
almost the same as that of the conventional single-channel TT, where the difference comes only from the cross-channel transformations.

\subsection{Three-stage training}
\vspace{-.5em}
\label{sec:three-stage}

Although VarArray is supposed to produce clean 
overlap-free signals,
the actual outputs contain different kinds of processing errors and artifacts. 
In particular, two types of errors, called \textit{leakage} and \textit{utterance split}, are unique to CSS and VarArray.
The leakage refers to generating the same speech content from both output channels. 
The utterance split refers to splitting one utterance into two or more pieces and generating them from different output channels. 
While these errors have a detrimental effect on existing systems~\cite{yoshioka2022vararray}, 
we can have the two-channel TT learn to handle such erroneous input.
For this purpose, 
we propose a three-stage training scheme as follows, which considers 
the fact that we have a limited amount of real multi-channel training data
while monaural single-talker ASR training data are available in large quantities. 
Below, let $S_0$ denote the large-scale single-talker ASR training set.


In the first stage,
a conventional (i.e. single-channel) t-SOT multi-talker TT is trained
by using a large-scale monaural multi-talker training set, $S_1$, which is derived from $S_0$ by simulation. 
By following \cite{kanda2021large}, 
each multi-talker training sample of $S_1$ can be 
generated by mixing
randomly chosen two single-talker audio samples from $S_0$ with a random delay.
During the model training, the samples are randomly picked from $S_0$ and $S_1$ at a 50:50 ratio.

In the second stage,
a two-channel TT is
initialized by copying the parameters of the t-SOT TT
trained in the first stage.
The parameters of the cross-channel transformation are randomly initialized.
The two-channel TT is then trained by using
a large-scale two-channel training dataset, $S_2$, which is obtained from $S_0$ by the following simulation method. 
(i) Randomly sample one utterance $u_1$ from $S_0$, and put it in the $c$-th channel with 
$c$ being randomly drawn from $\{0,1\}$.
(ii) With a 50\% chance, randomly sample another utterance $u_2$
from $S_0$, and put it in the $(1-c)$-th channel
after prepending a random delay of $d\sim\mathcal{U}(0, \mathrm{len}(u_1))$, where 
$\mathcal{U}$ is the uniform distribution.
(iii) Randomly split the generated two-channel audio into $p$-sec chunks,
where $p\sim\mathcal{U}(5,50)$,
 and swap the signals of the even-numbered chunks. 
 This simulates the utterance split phenomenon.
 (iv) For each channel, randomly chop the audio into  $q$-sec
  chunks, where $q\sim\mathcal{U}(1,5)$.  
  Then, for each channel of the even-numbered chunks, mix the audio signal of the other channel
  after scaling the volume by $v\sim\mathcal{U}(0, 0.2)$. 
  This step simulates the leakage phenomenon.
  
In the third stage,
the two-channel TT is further fine-tuned by using real multi-channel data.
VarArray is first applied to the multi-channel training samples
to generate two separated audio signals.
The two-channel TT from the second stage is then fine-tuned by using the two separated-signals and the corresponding transcription.
During the fine-tuning, the VarArray parameters are frozen.

\begin{table*}[t]
\ra{0.90}
\tabcolsep = 1.7mm
  \caption{WER (\%) for AMI development and evaluation sets with distant microphones. For B3 and B6, VarArray-based speech enhancement was applied by forming one MVDR beamformer generating monaural audio to perform only noise reduction without speech separation.}
  \label{tab:summary}
  \centering
{ \footnotesize
\begin{tabular}{@{}lrrrrllrrrrlllllll@{}}
    \toprule
ID & \multicolumn{4}{c}{Front-end configuration} && \multicolumn{5}{c}{Back-end configuration} && \multicolumn{3}{c}{Back-end training}  & \multirow{2}{*}{\shortstack[l]{Test\\segment}} & \multicolumn{2}{c}{WER (\%)}  \\ \cmidrule{2-5} \cmidrule{7-11} \cmidrule{13-15} \cmidrule{17-18} 
& In & Out & Param. & Latency && Model  & In & Cross & Param. &  Latency && 1ch-PT & 2ch-PT &  FT &  & dev & eval  \\ \midrule
B1 & - & - & - & - && Single-talker TT18 & 1 & - & 82M & 0.16 sec &&  75K & - & - & utt & 38.0 & 40.8   \\
B2 & - & - & - & - && Single-talker TT18 & 1 & - & 82M & 0.16 sec && 75K & - &  AMI & utt & 27.3 & 30.3   \\ 
B3 & 8 & 1 & 2M & 0.8 sec && Single-talker TT18 & 1 & - & 82M & 0.16 sec &&  75K & - &  AMI & utt& 25.8 & 27.9  \\  \hdashline[1pt/2pt]\hdashline[0pt/1pt]
B4 & - & - & - & - && t-SOT TT18 & 1 & - & 82M & 0.16 sec &&  75K-sim & - &  - & utt-gr& 35.5 & 40.3 \\
B5 & - & - & - & - && t-SOT TT18 & 1 & - & 82M & 0.16 sec &&  75K-sim & - &  AMI & utt-gr& 21.6 & 25.3 \\
B6 & 8 & 1 & 2M & 0.8 sec && t-SOT TT18 & 1 & - & 82M & 0.16 sec &&  75K-sim & - &  AMI & utt-gr& 20.7 & 23.0 \\ \midrule
P1 & 8 & 2 & 2M & 0.8 sec && t-SOT 2ch-TT18 & 2 & - & 82M & 0.16 sec &&  75K-sim &  - &   AMI & utt-gr& 19.3 & 21.7 \\  
P2 & 8 & 2 & 2M & 0.8 sec && t-SOT 2ch-TT18 & 2 & - & 82M & 0.16 sec &&  75K-sim &  75K-sim & AMI & utt-gr& 18.6 & 21.1 \\   
P3 & 8 & 2 & 2M & 0.8 sec && t-SOT 2ch-TT18 & 2 & Eq. \eqref{eq:scale} & 82M & 0.16 sec &&  75K-sim &  75K-sim & AMI & utt-gr& 18.5 & 21.0 \\   
P4 & 8 & 2 & 2M & 0.8 sec && t-SOT 2ch-TT18 & 2 & Eq. \eqref{eq:proj} & 84M & 0.16 sec &&  75K-sim &  75K-sim & AMI & utt-gr& 18.3 & 20.6 \\     \hdashline[1pt/2pt]\hdashline[0pt/1pt]
P5 & 8 & 2  & 2M & 0.8 sec && t-SOT 2ch-TT36 & 2 & Eq. \eqref{eq:proj}  & 142M & 0.64 sec &&  75K-sim &  75K-sim & AMI & utt-gr& 15.3 & 17.4 \\ 
P6 & 8 & 2 & 2M & 0.8 sec && t-SOT 2ch-TT36 & 2 & Eq. \eqref{eq:proj} & 142M & 2.56 sec &&  75K-sim &  75K-sim &  AMI & utt-gr&  14.4 & 16.5\\ 
P7 & 8 & 2  & 56M & 0.8 sec && t-SOT 2ch-TT36 & 2 & Eq. \eqref{eq:proj}  & 142M & 2.56 sec &&  75K-sim &  75K-sim &  AMI & utt-gr& 13.7  & 15.5 \\ \bottomrule
  \end{tabular}
  }
  \vspace{-5mm}
\end{table*}

\section{Experiments}
\label{sec:exp}
\vspace{-.5em}

\subsection{Data and metric}
\vspace{-.5em}

We used the AMI meeting corpus \cite{carletta2005ami}
for the fine-tuning and evaluation of the proposed system.
 The corpus contains approximately 100 hours of meeting recordings, 
 each containing three to five participants.
The audio
was recorded with an 8-ch microphone array.
We adopted the text formatting and data split of the
Kaldi toolkit \cite{povey2011kaldi}.
There are training, development and evaluation sets,
each of which contains 80.2 hr, 9.7 hr, and 9.1 hr of recordings.
The training set was used for the ASR model fine-tuning. The development and evaluation sets were used 
for the WER calculation.
We applied causal logarithmic-loop-based automatic gain control (AGC) to normalize the significant volume differences among different recordings. AGC was applied after the front-end processing.

In addition to the AMI corpus, we used 
64 million 
 anonymized and transcribed English utterances,
totaling 75K hours \cite{kanda2021large}, as $S_0$ 
for the first- and second-stage pre-training.
The data consist of monaural audio signals from various domains, such as voice command and dictation, 
and their reference transcriptions. 
 Each training sample is supposed to contain single-talker speech
 while it could occasionally contain untranscribed background human speech. 

In the evaluation, we followed the utterance and utterance-group segmentations proposed in 
\cite{kanda2021large}. 
The utterance group is defined as a set of adjacent utterances that
are connected by speaker overlap regions.
By following \cite{kanda2021large}, 
the utterance segmentation was used for the the single-talker ASR model evaluation 
while the utterance-group segmentation was used for the multi-talker ASR model evaluation.
For the WER calculation, 
we used the
multiple dimension Levenshtein edit distance calculation
implemented in 
ASCLITE \cite{fiscus2006multiple}.\footnote{To  reduce the computational complexity,
unlike the original way of using ASCLITE 
where the multiple references are aligned with a single sequence of time-ordered hypothesis, 
we calculated the WER by aligning the two hypothesis streams with a single sequence of time-ordered reference words obtained based on the official time stamps of the corpus.
This procedure allowed the WER to be calculated for all utterance-group regions within reasonable computation time.}

\subsection{Model configuration}
\vspace{-.5em}

\subsubsection{VarArray}
\vspace{-.5em}

VarArray consists of a set of multi-stream conformer blocks, a set of transform-average-concatenate (TAC) layers (optional)~\cite{luo2020tac}, a mean pooling layer, a set of single-stream conformer layers, and a mask prediction layer. The model input is a rank-3 tensor where the three axes correspond to  the channels, time frames, and features. The multi-stream conformer blocks are applied to each of the channel-wise slices of the input tensor for temporal modeling while the TAC layers are used to process each frame-wise slice for cross-channel modeling. They are interleaved with each other.
The mean pooling layer averages the features across the channels and is followed by the single-stream conformer layers for additional transformation. The mask prediction layer is a fully connected layer with softmax activation. See \cite{yoshioka2022vararray} for further details. 

Two VarArray models with different sizes were used to examine the impact that the front-end quality might have on the multi-talker ASR accuracy. They had 56M and 2M parameters with the latter targeted for resource-constrained edge processing. The bigger model had 3 multi-stream conformer blocks, where each block consisted of 
5 conformer layers, each with 8 attention heads, 512 dimensions, and 33 convolution kernels. The second block was sandwiched by 512-dim TAC layers. The model had 20 additional single-stream conformer layers having the same size as the multi-stream ones. The smaller model had one multi-stream block consisting of 3 conformer layers with 3 attention heads, 48 dimensions, and 33 convolution kernels, plus 8 additional single-stream conformer layers without TAC layers. It also performed 2x decimation and interpolation at the start and end of the model for processing cost reduction. Both models were trained on the 3K-hr simulated multi-channel data of \cite{yoshioka2022vararray}, where the number of microphones were randomly chosen between 3 and 7 for each mini-batch during training.

\subsubsection{t-SOT two-channel TT}
\vspace{-.5em}

We used a TT 
with chunk-wise look-ahead proposed in \cite{chen2021developing}.
The number of layers $L$ was set to 18 or 36.
We refer to the model with $L=18$ as TT18 and the model with
$L=36$ as TT36.
For the two-channel TT model, $N$ was set to 9 for TT18 while it was set to 12 for TT36
based on our preliminary experiment.
Each transformer block consisted of
a 512-dim multi-head attention with 8 heads 
and a 2048-dim point-wise feed-forward layer. 
The prediction network consisted of 
two layers of 1024-dim long short-term memory.
4,000 word pieces plus blank and $\langle \mathrm{cc}\rangle$ tokens
were used as the recognition units.
We used 80-dim log mel-filterbank extracted for every 10 msec.
As proposed in \cite{chen2021developing}, 
the algorithmic latency
of the model was controlled
based on the chunk size of the attention mask.

In the first-stage pre-training,
we performed 425K training iterations with
32 GPUs, each of which processed mini-batches of 24K frames.
We used 
a linear decay learning rate schedule with a peak learning rate 
of 1.5e-3 
after 25K warm up iterations.
In the second-stage pre-training,
we performed 100K training iterations with
8 GPUs, each of which processed mini-batches of 24K frames.
We used 
a linear decay learning rate schedule starting from a learning rate 
of 1.5e-4. 
Finally, in the fine-tuning stage,
we performed 2,500 training iterations with
8 GPUs, each processing mini-batches of 12K frames.
We used 
a linear decay learning rate schedule with an initial learning rate 
of 1.5e-4.

%

\subsection{Evaluation results}
\vspace{-.5em}
Table \ref{tab:summary} shows 
the WERs for various combinations of the VarArray-based front-end  
and back-end ASR configurations. 
The systems with IDs staring with B were built to clarify the contributions 
of individual configurations in a baseline setting. 
Systems B1 to B3 were based on single-talker ASR\footnote{In addition to B1-B3, we also evaluated a system that generated two separated signals with VarArray and then performed ASR for each separated signal with a single-talker model. However, due to the difficulty in fine-tuning the single-talker ASR model based on the separated signals (see Section \ref{sec:design}), this system produced 25.1\% and 29.2\% for the development and evaluation sets, respectively, which were on par or even worse than the other baselines.}.
System B4 to B6 used t-SOT for multi-talker ASR.
First, we can see that the AMI-based fine-tuning yielded 
significant WER improvements (B1$\rightarrow$B2, B4$\rightarrow$B5). Especially, the t-SOT TT18 obtained noticeably larger WER reductions, resulting in lower WERs 
than the single-talker TT18 (B2 vs. B5),
which was consistent with prior reports \cite{kanda2021large}.
Applying a single-output (i.e., traditional) multi-channel front-end, the WER was further improved from 25.3\% to 23.0\% for the evaluation set (B5$\rightarrow$B6).

We then evaluated the systems based on the proposed t-SOT-VA framework 
(P1--P7).
By comparing B6 and P1, we can see that the proposed 
two-channel TT achieved
a relative WER reduction of 5.7--6.8\% compared with the conventional single-input model.
Introducing the 2nd-stage pre-training further reduced the WERs by relative 2.8-3.6\% (P1$\rightarrow$P2).
Finally, the cross-channel transform (P3 and P4) further produced relative WER gains of 1.6--2.4\%.
Overall, compared with the baseline B6,
the proposed system achieved relative WER improvements of 
10.4--11.6\%.


Systems P5--P7 were evaluated to 
investigate the WER impact of the model size and latency.
As expected, the WERs were further improved by 
increasing the model size and the ASR latency.
The largest system, P7, achieved the WERs
of 13.7\% and 15.5\% for the development and evaluation sets, respectively.
To the best of our knowledge, these results represent the SOTA WERs 
for the AMI distant microphone setting by significantly 
outperforming previously reported results 
\cite{kanda2021large,zhang2022bigssl,wang2022leveraging}
while retaining the streaming inference capability.



Finally, Table \ref{tab:mic} shows 
 the WERs of P7 for different microphone numbers.
 We used the microphones
 indexed as [0], [0,1], [0,1,2], [0,2,4,6], [0,2,4,6,7], [0-3,5,6], [0-6], or [0-7]
  for each microphone number setting.
 The experimental result confirms that the proposed system can make use of different numbers and shapes of microphones without retraining.

\begin{table}[t]
\ra{0.9}
  \vspace{-3mm}
  \caption{WER (\%) of P7 system 
  applied for 2--8 microphones. Neither VarArray nor t-SOT models were retrained.
  Monaural t-SOT TT36 was applied for single-microphone case.}
  \label{tab:mic}
  \centering
{   \footnotesize
\begin{tabular}{@{}lcccccccc@{}}
    \toprule
\# of mics.    & 1 & 2 & 3 & 4 & 5 & 6 & 7 &  8   \\ \midrule
AMI-dev & 16.6   & 16.4 & 15.1 & 14.2 & 13.9 & 13.9 & 13.8 &  13.7 \\
AMI-eval & 19.7  & 18.8 & 17.2 & 16.4 & 15.9 & 15.8 & 15.7 & 15.5 \\ \bottomrule
  \end{tabular}
  }
  \vspace{-5mm}
\end{table}




\section{Conclusion}
\vspace{-.5em}
This paper proposed t-SOT-VA, a novel streaming DCSR framework
for arbitrary microphone arrays,
 by integrating  VarArray and t-SOT-based multi-talker ASR.
To achieve the best combination of both technologies under practical settings, 
we designed a t-SOT-based ASR model that takes in two separated speech signals from VarArray. 
The evaluation 
using the AMI meeting corpus 
 showed that the system based on the proposed framework significantly outperformed conventional DCSR systems
while retaining the streaming inference capability. 

\bibliographystyle{IEEEtran}
\bibliography{mybib}

\newpage
\appendix
\section{Impact of speech overlaps and speaker-to-microphone distance}

Table \ref{tab:ihm} shows the WERs for different types of input signals. 
This experiment was carried out to analyze the impacts of speaker-to-microphone distances and speech overlaps.
In this experiment,
we used TT36 with 2.56 sec latency by using the most appropriate training and testing framework for each input type.
The first row shows the WERs obtained with 
the independent headset microphone (IHM) signals and 
a single-talker TT36, where the model was pre-trained on the 75K-hour data and fine-tuned on the AMI IHM training data. 
The second row shows the WERs obtained by using the
mixture of the IHM signals (a.k.a. IHM-MIX) as input, 
where we used a t-SOT TT36 that was pre-trained on the 75K-hour data based simulation data and fine-tuned on the AMI IHM-MIX training data.
Finally, the third row shows the WERs of our P7 system applied to the multiple distant microphone (MDM) signals.

By comparing the first and second rows, 
we can see a 32--35\% WER increase due to the speech overlaps. 
This degradation is attributed partly to our setting of $M=2$ , which inevitably degraded the WER for speech regions where more than two speakers overlap.
Comparing the second and third rows shows 
a further 14--16\% degradation due to the speaker-to-microphone distance.
Overall, we still observed noticeable WER degradations resulting from the speech overlaps and speaker-to-microphone distances, which calls for further technology improvement.



\begin{table}[h]
\ra{0.9}
\tabcolsep = 1mm
  \caption{WERs (\%) for various input types. TT36 with 2.56 sec latency was used with the most appropriate training and testing framework for each input type.}
  \label{tab:ihm}
  \centering
{   \footnotesize
\begin{tabular}{@{}lllllrr@{}}
    \toprule
Model  & Input &  \multirow{2}{*}{\shortstack[l]{Test\\segment}}  &  \multirow{2}{*}{\shortstack[l]{Speech\\overlaps}}  &  \multirow{2}{*}{\shortstack[l]{Spk-to-mic.\\distance}}  & \multicolumn{2}{c}{WER (\%)}   \\  \cmidrule{6-7}
          &      &  &  &   & dev & eval   \\ \midrule
Single-talker TT36 & IHM & utt & no & close  &  9.1 & 9.9  \\
t-SOT TT36 & IHM-MIX & utt-gr & yes & close   & 12.0   &  13.4 \\
t-SOT-VA TT36 & MDM  & utt-gr & yes & distant  & 13.7 & 15.5  \\ \bottomrule
  \end{tabular}
  }
\end{table}

\end{document}